\documentclass[conference]{IEEEtran}
\IEEEoverridecommandlockouts
\PassOptionsToPackage{hyphens}{url}
\usepackage[T1]{fontenc}
\usepackage[utf8]{inputenc} 
\DeclareUnicodeCharacter{00A0}{~}   
\DeclareUnicodeCharacter{2009}{\,}  
\DeclareUnicodeCharacter{202F}{\,}  
\usepackage{lmodern}
\usepackage{microtype}
\usepackage{cite}
\usepackage{graphicx}
\usepackage{subcaption}
\usepackage{booktabs}
\usepackage{tabularx}
\usepackage{amsmath,amssymb}
\usepackage[inline]{enumitem}
\usepackage{listings}
\usepackage{xcolor}  
\lstdefinelanguage{JavaScript}{
  keywords={typeof, new, true, false, catch, function, return, null, 
    switch, var, if, in, while, do, else, case, break, for, let, const},
  keywordstyle=\color{blue}\bfseries,
  ndkeywords={class, export, boolean, throw, implements, import, this},
  ndkeywordstyle=\color{purple}\bfseries,
  identifierstyle=\color{black},
  sensitive=true,
  comment=[l]{//},
  morecomment=[s]{/*}{*/},
  commentstyle=\color{gray}\ttfamily,
  stringstyle=\color{red}\ttfamily,
  morestring=[b]',
  morestring=[b]"
}

\lstdefinelanguage{TypeScript}[]{JavaScript}{
  morekeywords={
    abstract, as, asserts, any, unknown, never, readonly,
    private, public, protected, enum, interface, type,
    implements, declare, namespace, module, require,
    constructor, super
  }
}
\lstdefinelanguage{Solidity}{
  morekeywords={
    pragma,solidity,contract,library,using,struct,enum,event,modifier,
    function,constructor,returns,return,emit,revert,require,assert,
    view,pure,payable,nonpayable,public,private,internal,external,
    virtual,override,constant,immutable,unchecked,assembly,new,delete,
    if,else,for,while,do,break,continue,mapping,
    address,bool,string,bytes,bytes1,bytes4,bytes32,
    int,int8,int16,int32,int64,int128,int256,
    uint,uint8,uint16,uint32,uint64,uint128,uint256
  },
  sensitive=true,
  comment=[l]{//},
  morecomment=[s]{/*}{*/},
  morestring=[b]",
  morestring=[b]',
  showstringspaces=false,
  keywordstyle=\bfseries,
  stringstyle=\ttfamily,
  commentstyle=\ttfamily\itshape,
  columns=fullflexible,
  upquote=true
}
\usepackage{titlesec}
\usepackage{longtable}

\usepackage{hyperref}
\hypersetup{
  colorlinks,
  linkcolor=black,
  citecolor={blue!60!black},
  urlcolor ={blue!60!black}
}

\lstdefinelanguage{json}{
    basicstyle=\ttfamily\scriptsize,
    showstringspaces=false,
    breaklines=true,
    frame=single,
    backgroundcolor=\color{gray!5},
    literate=
     *{0}{{{\color{blue}0}}}{1}
      {1}{{{\color{blue}1}}}{1}
      {2}{{{\color{blue}2}}}{1}
      {3}{{{\color{blue}3}}}{1}
      {4}{{{\color{blue}4}}}{1}
      {5}{{{\color{blue}5}}}{1}
      {6}{{{\color{blue}6}}}{1}
      {7}{{{\color{blue}7}}}{1}
      {8}{{{\color{blue}8}}}{1}
      {9}{{{\color{blue}9}}}{1}
      {:}{{{\color{black}:}}}{1}
      {,}{{{\color{black},}}}{1}
      {"}{{{\color{black}"}}}{1}
}

\titlespacing*{\section}{0pt}{*0.9}{*0.6}   
\titlespacing*{\subsection}{0pt}{*0.7}{*0.5}
\hypersetup{
    colorlinks,
    linkcolor={black},
    citecolor={blue!60!black},
    urlcolor ={blue!60!black}
}

\IEEEaftertitletext{\vspace{-1.2\baselineskip}}

\title{%
A GAIA-X-Aligned, Blockchain-Anchored\\
Privacy-Preserving, Zero-Knowledge Digital Passport\\
for Smart Vehicles%
}

\author{\IEEEauthorblockN{Pradyumna Kaushal}
\IEEEauthorblockA{Dept.\ of Production Engineering, NIT Tiruchirappalli, India\\
\texttt{kaushal.pradyumna@gmail.com}}}

\begin{document}
\maketitle

\begin{abstract}
\textbf{Problem-}From the OEM production line through successive owners, a vehicle accumulates a rich yet fragmented life-cycle record that today is stored in opaque silos, cannot be independently verified, and forces insurers, service centres, regulators, and buyers into costly \emph{Know-Your-Customer} (KYC) procedures.

\textbf{Contribution-}\emph{VehiclePassport} introduces a trust-less, privacy-preserving digital passport whose on-chain hash immutably commits to \emph{every event}-manufacturing, maintenance, ownership transfers, telemetry and internal (battery,motor) health-while zero-knowledge (ZK) proofs let stakeholders validate high-level statements without seeing raw data.  A short-lived, field-scoped JWT conveys a selective JSON-LD credential; Polygon zkEVM ensures anchored logs cannot be altered; and GAIA-X self-descriptions enable sovereign, cross-border interoperability.

\textbf{Results-}An open-source reference stack (backend,API's, SDK and telemetry middleware) processes 1 Hz streams for a million vehicles, anchors hashes for \$0.02 each, and validates ZK proofs in sub-10 ms-empowering insurers, service centres, resellers and owners to transact instantly with cryptographic assurance.

\textbf{Impact-}The passport collapses paperwork, removes centralized trust, and establishes an auditable continuum from factory provisioning to the current driver, laying the groundwork for global, GDPR-compliant mobility data markets.
\end{abstract}

\section{\textbf{Introduction}}
Modern vehicles are secure IoT datacentres on wheels, emitting up to \(\sim\)25 GB h\(^{-1}\) of diagnostics and telemetry.  Yet resale buyers, insurers and regulators still rely on paper service books or unverifiable PDFs.  We propose \emph{VehiclePassport}, a cryptographically verifiable passport in which

\begin{enumerate*}[label=(\roman*)]
\item every vehicle is issued a \emph{vehicle-bound (soul-bound)} ERC-721 token whose ownership can transfer between people while remaining permanently attached to the physical car,
\item that token points to an immutable on-chain hash of the vehicle’s JSON-LD life-cycle document,
\item selective ZK proofs replace hand-written bills of sale, RTO paperwork and KYC checks, and
\item GAIA-X metadata makes the passport discoverable and machine-interpretable across EU data spaces.
\end{enumerate*}

In short, the NFT declares \emph{who} currently owns the asset, while the passport declares \emph{everything you need to know or prove} about that asset-without leaking private telemetry.

\section{\textbf{Background and Related Work}}
\subsection{Legacy Vehicle-Data Ecosystems}
Pre-connected vehicles funnel telemetry into proprietary OEM clouds or national databases.  These central stores cannot
\begin{enumerate*}[label=(\alph*)]
\item give a buyer cryptographic confidence in battery or odometer claims,
\item let a driver prove \emph{good driving behaviour} to an insurer without revealing routes, or
\item stop an insider from editing records after the fact.
\end{enumerate*}

\subsection{Need for Privacy-Preserving Selective-Disclosure}
A privacy-preserving vehicle passport solves three long-standing problems:

\begin{itemize}[leftmargin=*]
\item \textbf{Trust gap} - buyers distrust owner or OEM claims; ZK lets the car \emph{prove} it meets a policy.
\item \textbf{One-size-fits-all data dumps} - regulators need CO\(_2\) ranges, insurers need mileage, but no one needs raw GPS traces.
\item \textbf{Mutability} - post-event edits in centralized DBs enable odometer fraud; chained hashes prevent it.
\end{itemize}

\subsection{Blockchain Approaches}
Projects such as Bosch AMS or early MOBI VID store whole JSON payloads on-chain, inflating gas cost and breaching GDPR.  Others mint transferrable VIN NFTs but lack selective disclosure.  No prior work merges a soul-bound vehicle NFT, off-chain JSON-LD, ZK proofs \emph{and} GAIA-X compliance.

\subsection{GAIA-X Primer}
GAIA-X (v3.0) mandates machine-readable \textbf{self-descriptions} in JSON-LD, W3C \textbf{verifiable credentials}, and policy endpoints for federated usage control.  VehiclePassport realises these artefacts (Sec.~\ref{sec:gaia}) and effectively becomes a decentralized, standards-based identifier that any EU data space can index.

\section{\textbf{System Goals}}
\begin{enumerate}[label=\textbf{G\arabic*},leftmargin=*]
\item \textbf{Integrity}.  Off-chain artefacts are SHA-256 hashed, on-chain identifiers use Keccak-256; every digest is anchored so any tampering becomes publicly detectable.
\item \textbf{Selective Privacy}.  Short-lived, field-scoped JWTs deliver only the data the requester is entitled to, while Groth16 ZK-proofs attest to those facts against the owner-held telemetry.  
      A verifier recomputes the on-chain hash, validates the proof, and-when everything matches-shows a \emph{“Verified-by-Matter ZK™”} badge.  Raw sensor data stays local; nothing sensitive ever leaves the owner’s domain.

\item \textbf{GAIA-X Alignment}.  Passport stored as JSON-LD; policy API and verifiable credentials satisfy GAIA-X meta-model.
\item \textbf{Economic Viability}.  Hash anchoring costs \(\le\)\$0.02 on Polygon zkEVM, and the same contracts port unchanged to any Ethereum-equivalent zk-roll-up (StarkNet, Scroll, Optimism Bedrock, etc.), keeping fees low and throughput high.
\item \textbf{Trust-less Transfer}.  Ownership moves through a dual-party signature handshake: both seller and buyer co-sign the sale record \(\rightarrow\) hash is anchored on-chain \(\rightarrow\) the vehicle-bound NFT is safely transferred-eliminating paper RTO / KYC friction.

\item \textbf{User Flow}.  
\begin{enumerate*}[label=(\alph*)]
\item Vehicle registered → JSON-LD in Postgres,
\item hash → blockchain,
\item GAIA-X self-description + access-policy endpoints exposed.
\end{enumerate*}
\end{enumerate}

\section{\textbf{Architecture Overview}}
\begin{figure}[ht]
  \centering
  \includegraphics[width=\linewidth]{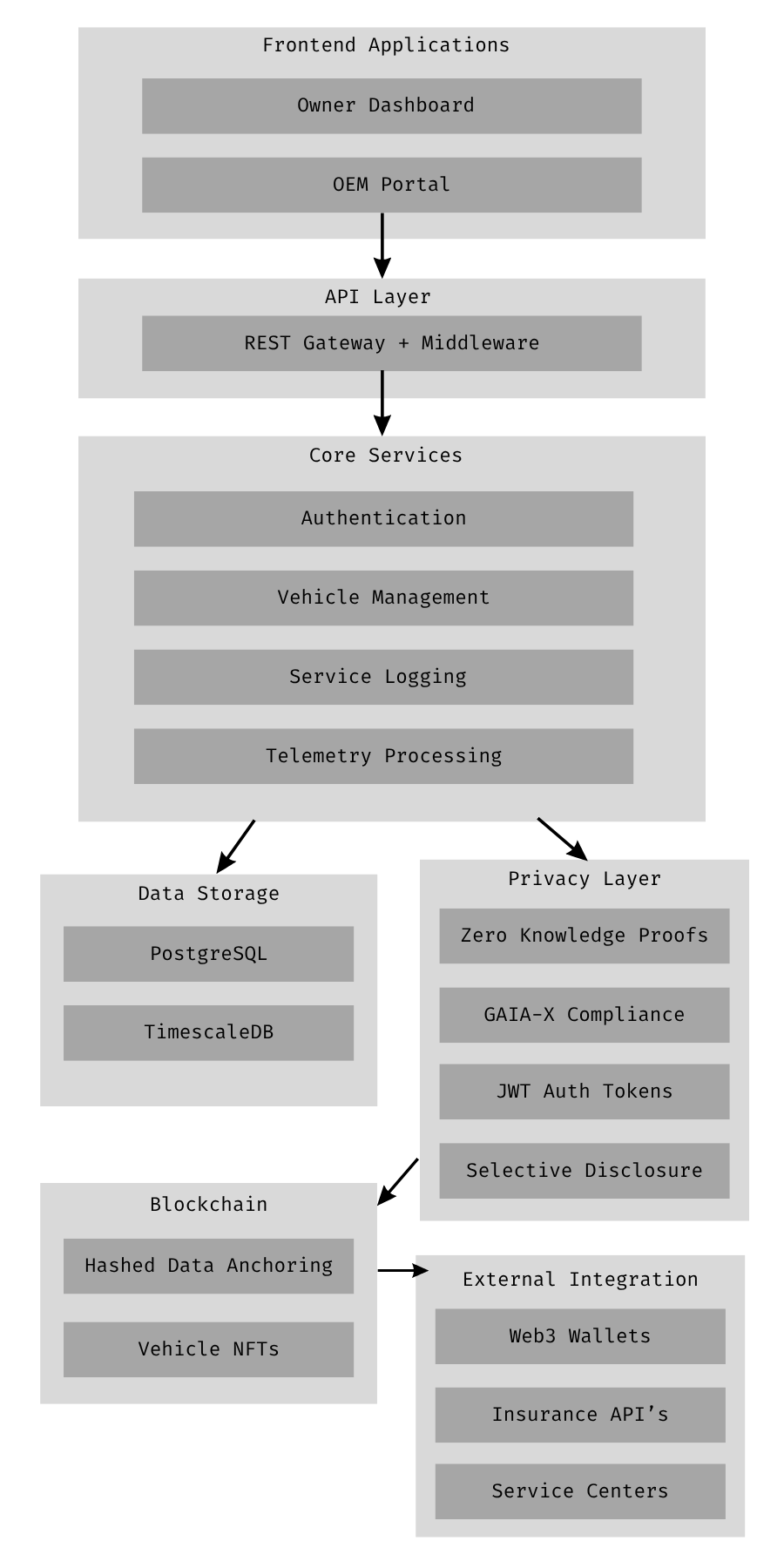}
  \caption{Layered Architecture: manufacturing seed $\rightarrow$ telemetry ingest $\rightarrow$ zkEVM-anchor $\rightarrow$ SDK's and APIs.}
  \label{fig:arch}
\end{figure}

Fig.~\ref{fig:arch} abstracts the stack into five canonical layers.  
To guide the reader from that \emph{bird-eye view} to the deep-dive
subsections that follow, we first summarise the concrete artefacts that
instantiate each layer in the running implementation.

\subsubsection{\textbf{Front-end Layer}}\label{sec:frontend}

The platform ships five thin-client React applications, all backed by the
same REST/GraphQL gateway and a shared TypeScript SDK (vehicle-nft~\& zk-api
wrappers).  Users authenticate by signing an \emph{EIP-4361 “Sign-in with
Ethereum”} challenge from any ERC-4337 wallet; the session cookie then carries
a short-lived JWT.

\begin{description}[leftmargin=*]
\item[Owner Dashboard ]
  A single-page cockpit where a private owner
  \begin{enumerate*}[label=(\alph*)]
    \item views every vehicle passport with
          \emph{ZK-Verified, On-Chain} badges,
    \item approves or denies \texttt{/api/access-request} tickets,
    \item signs pending service logs before they are anchored,
    \item reviews the immutable service history,
    \item and launches \emph{Sell / Buy} wizards that call the NFT-SDK
          to generate the dual-signature transfer payload.
  \end{enumerate*}

\item[Access Request Portal.]
  A public form for insurers, workshops or regulators:
  scan the VIN QR, tick required fields, submit-behind the scenes the page
  POSTs \texttt{/api/access-request}.  When the owner later approves, the
  requester pastes the issued JWT into the “Verify Token” tab to retrieve only
  the whitelisted JSON-LD attributes.

\item[Service Centre Portal.]
  Lets a workshop push a signed maintenance
  description; the owner’s dashboard shows it under \emph{Pending Logs} until
  they countersign, after which the backend calls
  \texttt{ServiceLogAnchor.sol}.

\item[Matter Production Portal.]
  A semi-centralised UI for OEM staff to
  \textit{create} production vehicles, mint the initial NFT, and hand the
  record to point-of-sale clerks who specify the buyer’s wallet.  This is the
  only place where custodial keys are used; everything downstream is fully
  decentralised.

\item[Insurance ZK Playground.]
  A demo site that calls the
  \texttt{/api/zkp/\{generate,verify\}} routes so actuaries can ask questions
  such as “batteryHealth~$\geq$~85\,\% \& mileage $<25$\,000 km” and verify the
  Groth16 proof completely client-side-no raw telemetry ever leaves the
  owner’s browser.
\end{description}

All five front-ends share
\texttt{@vehicle-passport/sdk}: a 2-kLOC library that
\begin{enumerate*}[label=(\roman*)]
  \item packs JSON-LD + pubkey with \texttt{ethers.solidityPacked} and
        computes the Keccak-256 to mirror the on-chain \texttt{bytes32},
  \item wraps NFT mint/transfer calls with automatic Gasless ERC-2771 relay,
  \item exposes helpers for fetching\,/\,verifying scoped JWTs and ZK proofs.
\end{enumerate*}

\subsubsection{\textbf{API Gateways \& Public Routes}}\label{sec:api}

The backend is fronted by two lightweight Express services that share
the same JWT / signature helpers but run in separate containers:
\begin{enumerate}
\item \textbf{Main API}. Mounts \texttt{/api/*}
      (\emph{access, service-log, owner, matter, sync-telemetry})
      plus the public \texttt{/vehicle/*} routes.
      It backs the Owner Dashboard and Matter Portal.

\item \textbf{ZKP API}. Exposes \texttt{/api/zkp/*}
      (proof \emph{generate, verify, anchor, status})
      and \texttt{/api/vkey/*}. Deploying it in an isolated pod keeps
      large circuit artefacts and witness generation off the
      critical-request path.
\end{enumerate}

Every request passes through the same middleware stack:  
\emph{CORS} $\rightarrow$ JSON body-parser $\rightarrow$
JWT/Keccak-sig verifier $\rightarrow$ route handler.
Tables \ref{tab:access-routes}–\ref{tab:zkp-routes} give an OpenAPI-level
overview; all payloads are \texttt{application/json}.

\setlength{\tabcolsep}{3pt}   

\begin{table}[ht]
\caption{Selective-Disclosure and Ownership Routes}
\scriptsize\centering
\setlength{\tabcolsep}{3pt}     
\begin{tabularx}{\linewidth}{@{}l l X@{}}
\toprule
\textbf{HTTP} & \textbf{Path} & \textbf{Purpose / Notes}\\\midrule
POST & \texttt{/access-request}            & Create request for \emph{field-scoped} access; body includes \{\textit{vehicleId, requester, fields[]}\}.\\
POST & \texttt{/approve/:id}               & Owner signs approval, gateway issues one-time JWT.\\
GET  & \texttt{/access/:token}             & Redeem token $\rightarrow$ minimised JSON-LD $+$ “Verified by Matter ZK™”.\\
GET  & \texttt{/owner/approvals}           & Owner dashboard: pending requests.\\
\addlinespace
GET  & \texttt{/vehicle/owner/:wallet}     & Portfolio lookup (VINs, anchors, NFT-ids).\\
POST & \texttt{/vehicle/initiate-transfer} & Dual-party off-chain payload (seller sig).\\
POST & \texttt{/vehicle/confirm-transfer}  & Buyer co-signs; backend calls \texttt{safeTransferFrom}, then anchors hash.\\
\bottomrule
\end{tabularx}
\label{tab:access-routes}
\end{table}

\begin{table}[ht]
\caption{Service-Log and Telemetry End-points}
\scriptsize\centering
\setlength{\tabcolsep}{3pt}
\begin{tabularx}{\linewidth}{@{}l l X@{}}
\toprule
\textbf{HTTP} & \textbf{Path} & \textbf{Purpose / Notes}\\\midrule
POST & \texttt{/service-log/request}             & Workshop submits log (hash $+$ sig) $\rightarrow$ \textit{pending}.\\
GET  & \texttt{/service-log/pending/:ownerId}    & Owner reviews / signs.\\
POST & \texttt{/service-log/approve/:id}         & Anchor to \texttt{ServiceLogAnchor.sol}; returns tx-hash.\\
\addlinespace
POST & \texttt{/sync-telemetry/:vehicle}         & 1 Hz or batch ingest; \texttt{x-api-key} required.\\
GET  & \texttt{/sync-telemetry/latest/:vehicle}  & Latest row for dashboards and ZK inputs.\\
\bottomrule
\end{tabularx}
\label{tab:service-routes}
\end{table}

\begin{table}[ht]
\caption{Matter OEM Portal / Production Routes}
\scriptsize\centering
\setlength{\tabcolsep}{3pt}
\begin{tabularx}{\linewidth}{@{}l l X@{}}
\toprule
\textbf{HTTP} & \textbf{Path} & \textbf{Purpose / Notes}\\\midrule
GET  & \texttt{/matter/vehicles}     & Inventory grouped by \emph{production} / \emph{customer}.\\
POST & \texttt{/matter/vehicles}     & Create production vehicle, compute hash, anchor, mint NFT.\\
POST & \texttt{/matter/sell-vehicle} & Point-of-sale flow; streams transfer status events.\\
POST & \texttt{/matter/mint-nft}     & (Re)mint NFT when legacy stock is imported.\\
\bottomrule
\end{tabularx}
\label{tab:matter-routes}
\end{table}

\begin{table*}[ht]
\caption{Zero-Knowledge Proof (ZKP) API End-points}
\scriptsize\centering
\setlength{\tabcolsep}{4pt}
\begin{tabularx}{\textwidth}{@{}l l l l X@{}}
\toprule
\textbf{Group} & \textbf{HTTP} & \textbf{Path} & \textbf{Body Keys} & \textbf{Outcome}\\\midrule
\multicolumn{5}{@{}l}{\textit{Proof-generation routes}}\\\midrule
– & POST & \texttt{/zkp/batteryHealth}      & vehicleId, threshold  & Proves $\textit{batteryHealth}>\theta$ without revealing the raw value (Groth16).\\
– & POST & \texttt{/zkp/mileage}            & vehicleId, threshold  & Proves $\textit{mileage}<\theta$.\\
– & POST & \texttt{/zkp/warrantyExpiry}     & vehicleId, timestamp  & Proves warranty valid beyond timestamp.\\
– & POST & \texttt{/zkp/accessRequestCount} & vehicleId, threshold  & Proves \#accessRequests $\le\theta$.\\
– & POST & \texttt{/zkp/serviceLogCount}    & vehicleId, threshold  & Proves \#serviceLogs $\ge\theta$.\\
\midrule
\multicolumn{5}{@{}l}{\textit{Verification-key routes}}\\\midrule
– & GET  & \texttt{/vkey/batteryHealth}      & – & Verification key for \texttt{batteryHealth} circuit.\\
– & GET  & \texttt{/vkey/mileage}            & – & Key for \texttt{mileage} circuit.\\
– & GET  & \texttt{/vkey/warrantyExpiry}     & – & Key for \texttt{warrantyExpiry}.\\
– & GET  & \texttt{/vkey/accessRequestCount} & – & Key for \texttt{accessRequestCount}.\\
– & GET  & \texttt{/vkey/serviceLogCount}    & – & Key for \texttt{serviceLogCount}.\\
\bottomrule
\end{tabularx}
\label{tab:zkp-routes}
\end{table*}

\textbf{Security defaults}.  
Every mutating request must be signed with a \texttt{Keccak-256}S wallet signature and passes through a common \texttt{JWT/Auth} middleware layer that enforces role-based scopes, rate limits, and least-privilege \texttt{JWTs}.  All state-changing calls emit on-chain receipts, giving an end-to-end, fully auditable trail.

\subsubsection{\textbf{Core Micro-services Layer}}\label{sec:core-services}

The backend is decomposed into five stateless containers; Each pod
shares the same \textit{JWT \(\rightarrow\) Keccak-sig} middleware but owns a
separate route group and Prisma schema.  State lives either in the
\texttt{vehicle} / \texttt{owner} tables (PostgreSQL) or in the
\texttt{telemetry\_hypertable} (TimescaleDB); every mutating route emits an
on-chain anchor or an audit-row so the flow remains end-to-end verifiable.

\subsubsection*{\textbf{A.\;Wallet-based Authentication}}\label{svc:auth}
\begin{itemize}[leftmargin=*]
\item \textbf{Passwordless wallet login.}  
      Pressing “Connect Wallet’’ triggers the standard
      \texttt{eth\_requestAccounts} flow (via \textsf{ethers.js}).  
      The returned \texttt{0x…} address is looked-up in the
      \texttt{Owner} table; if a row exists, that owner is considered
      “authenticated’’ and receives a 30-min JWT scoped to the
      primary-key \texttt{wallet}.  No e-mail or password ever crosses the
      wire-possession of the private key is the only credential.

\item \textbf{Ownership mapping.} On first login the service creates an
      \texttt{Owner} row  
      \(\langle id,\;wallet,\;displayName,\;email \rangle\) and
      thereafter resolves all business calls via the wallet address;
      multi-sig or ERC-4337 smart-wallets are equally supported.
\item \textbf{Matter Portal.} The production team uses the same flow; their
      wallet is merely flagged \texttt{role = "OEM"} which unlocks the
      additional \texttt{/matter/*} routes.
\end{itemize}

\subsubsection*{\textbf{B.\;Vehicle Management \& Passport Consistency}}\label{svc:veh-mgr}

Every vehicle row keeps a \emph{canonical hash} that must equal the anchor
stored in \texttt{VehiclePassportAnchor.sol}.  A mismatch surfaces as a
“\textit{Out-of-date hash}” badge in the dashboard, prompting the owner to
re-anchor or investigate tampering.

\paragraph*{GAIA-X self-description helper.}
When the row is first inserted (or whenever a mutable field changes) the SDK
generates a GAIA-compliant JSON-LD credential and its SHA-256 digest; the
latter is what is actually written on-chain.

\begin{lstlisting}[language=TypeScript,
caption=Issuer-side helper for GAIA-X Vehicle Self-Description,
label=lst:gaia-helper]
import crypto from "crypto";

export function generateGaiaCredential(v: any) {
  return {
    "@context": "https://www.w3.org/ns/credentials/v2",
    id: `urn:matter:vehicle:${v.vin}`,
    type: ["VerifiableCredential", "VehiclePassport"],
    issuer: "did:web:matter.in",
    issuanceDate: new Date(v.createdAt).toISOString(),
    credentialSubject: {
      vehicleVIN: v.vin,
      model: v.model,
      manufacturer: v.manufacturer,
      batteryHealth: v.batteryHealth,
      mileage: v.mileage,
      warrantyValidUntil: v.warrantyExpiry
    }
  };
}

export function hashGaiaCredential(cred: object): string {
  return crypto.createHash("sha256")
               .update(JSON.stringify(cred))
               .digest("hex");
}
\end{lstlisting}

\noindent
\noindent
Every \texttt{VehicleNFT} stores a 32-byte commitment hash directly inside the
token’s storage slot via the \texttt{vehicleDataHash[vehicleId]} mapping. This hash-typically a \texttt{keccak256(pubkey‖JSON-LD)}-
is passed during minting and acts as a constant-size fingerprint of the
vehicle’s off-chain passport.

\smallskip

The same hash is later re-anchored on-chain through
\texttt{VehiclePassportAnchor.anchor()}, ensuring cryptographic consistency
across layers. Together, this design allows any verifier to:

\begin{enumerate*}[label=(\alph*)]
\item fetch the on-chain \texttt{vehicleDataHash} from the NFT via a single
      \texttt{eth\_call}, 
\item recompute the hash locally using the owner’s public key and JSON-LD, and
\item verify passport integrity in \(O(1)\) time without exposing raw data.
\end{enumerate*}

This constant-time hash check enables scalable audits, privacy-preserving
ZKPs, and tamper-evident ownership without revealing any sensitive vehicle
information on-chain.

\subsubsection*{C.\textbf{\;Selective Disclosure Service}}\label{svc:sel-disc}

\begin{figure}[ht]
  \centering\includegraphics[width=\linewidth]{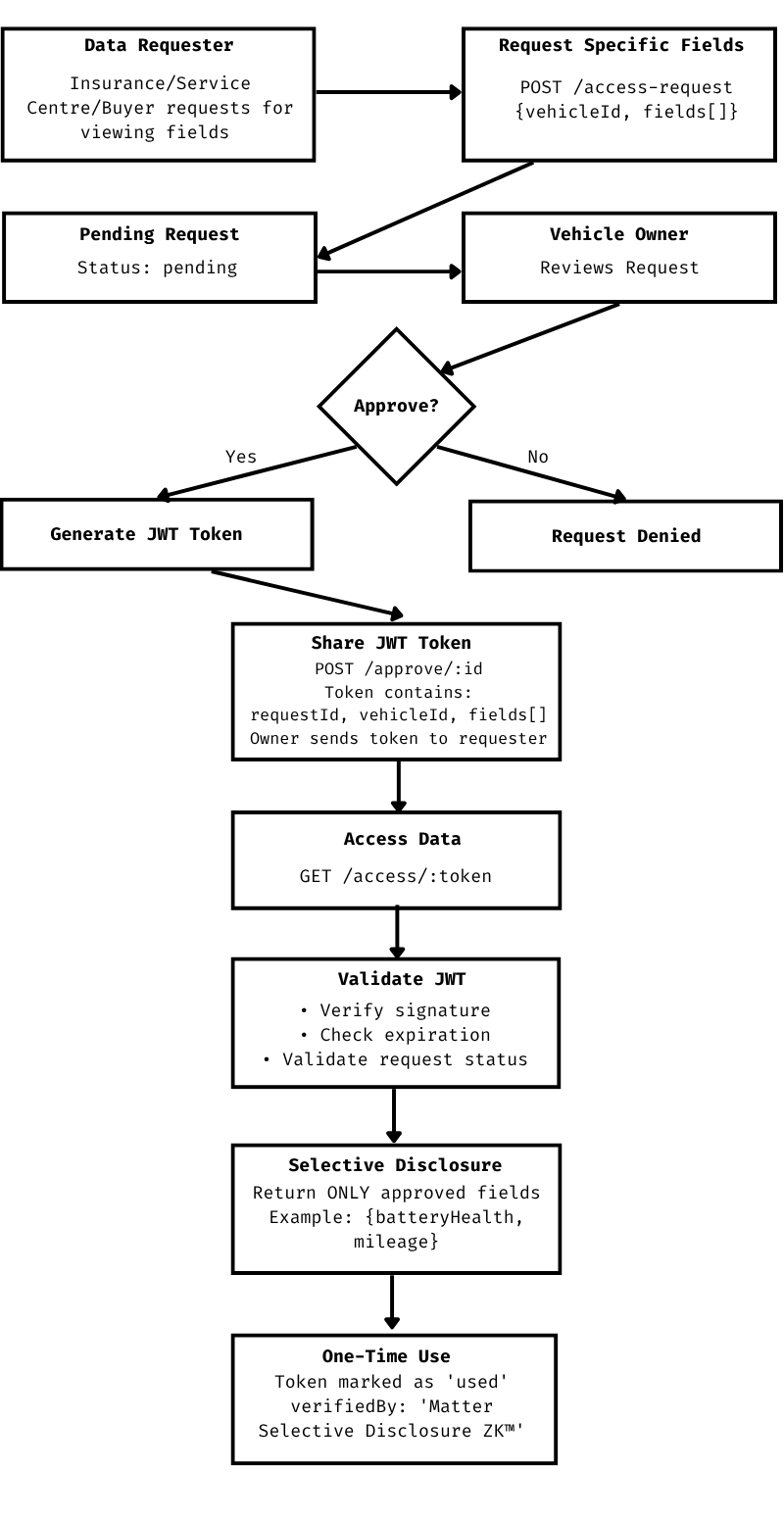}
  \caption{JWT Selective Disclosure Flow}
  \label{fig:jwt-flow}

\end{figure}
This service turns ownership into field-level data sharing while guaranteeing
privacy via \emph{scoped, one-time JWTs}.  Figure \ref{fig:jwt-flow} shows the
happy path; the text below annotates the three phases.

\paragraph*{\textbf{Phase 1 - request.}}
A third-party caller submits:

\smallskip
\centerline{\texttt{POST /api/access-request}}
\smallskip

\noindent
with body
\(\langle \textit{vehicleId},\; \textit{requester},\;
\textit{fields[]} \rangle\).
Library validation rejects any field not in the hard whitelist
\{\texttt{batteryHealth,\allowbreak mileage,\allowbreak warrantyExpiry,\allowbreak model,\allowbreak manufacturer}\}
. A DB row
is created with status=\textsc{pending} and a 10 min expiry.

\paragraph*{\textbf{Phase 2 - owner approval.}}
The Owner Dashboard pulls pending rows via
\texttt{GET /owner/approvals}.  
When the user clicks \emph{Approve} the backend signs and returns the
token in Listing \ref{lst:scoped-jwt}; the DB status flips to
\textsc{approved}.

\begin{lstlisting}[language=json, caption={One-time, field-scoped JWT}, label={lst:scoped-jwt}]
{
  "requestId": "e1c7",
  "vehicleId": "42",
  "fields": ["batteryHealth", "warrantyExpiry"],
  "iat": 1733097600,
  "exp": 1733098200
}
\end{lstlisting}

\paragraph*{\textbf{Phase 3 - redemption.}}
The requester redeems the token with
\texttt{GET /api/access/\{token\}}.
Middleware verifies the HMAC → DB row, enforces \texttt{exp},
projects only the approved columns, returns the JSON and finally marks the
row \textsc{used}.  Any replay attempt now yields \texttt{HTTP 410 Gone}.

\paragraph*{\textbf{Security posture.}}
\begin{itemize}[leftmargin=*]
\item \textbf{One-time use} stops replay attacks;
\item \textbf{10 min TTL} on both the pending row and the JWT bounds the
      exposure window;
\item \textbf{Signature header} (\texttt{X-Sig-Keccak}) for every mutating
      route ties  database writes to an EOA/private-key event;
\item \textbf{Comprehensive audit trail} (\textsc{pending}→\textsc{approved}
      →\textsc{used}) supports GDPR Art.~30 logging.
\end{itemize}


\subsubsection*{D. \textbf{\;Service History Logging}}\label{svc:sev-his}
This module enables cryptographically signed and owner-approved service records that are permanently attached to a vehicle's passport and anchored on-chain.

\textit{\textbf{Phase 1 - Service Center Initiation.}}  
A service center begins by scanning the QR code on the vehicle or manually entering the vehicle ID. They then input a detailed service description and connect their wallet ( service centre private key ) to submit the log.

\textbf{Portal Flow:}  
On the Service Portal, the center submits their data: \texttt{vehicleId}, \texttt{description}, and \texttt{centerEmail}. The system creates a deterministic JSON payload, hashes it using \texttt{keccak256}, and generates a digital signature using the center’s private key via MetaMask.

\textbf{Security Basis:}  
The signature proves:
\begin{itemize}
  \item The service center owns the private key (identity).
  \item They authorized the specific service log (intent).
\end{itemize}

\textit{\textbf{Phase 2 - Backend Validation.}}  
The backend verifies the signature by recovering the address and comparing it to the submitted one. If valid, it stores a \texttt{PendingServiceLog} with status=\textsc{pending} and links it to the vehicle.

\textit{\textbf{Phase 3 - Owner Review.}}  
The vehicle owner dashboard fetches all pending logs via database joins. Each log displays the description, center email, and original hash. The owner physically verifies the work done and proceeds to sign the same hash using MetaMask.

\textbf{Dual-Signature Security:}  

Both signatures bind to the same hash, proving mutual consent. The service center can’t deny they performed the service, and the owner can’t deny their approval, as illustrated in Fig.~\ref{fig:dual-sig}.

\textit{\textbf{Phase 4 - Blockchain Anchoring.}}  
Once both signatures are collected, the backend calls the \texttt{ServiceLogAnchor} smart contract on the Polygon testnet. The contract:
\begin{itemize}
  \item Rejects duplicate hashes.
  \item Emits a \texttt{LogAnchored} event.
  \item Stores the hash immutably on-chain.
\end{itemize}

\textit{\textbf{Phase 5 - Finalization.}}  
The system creates a finalized \texttt{ServiceLog} record containing:
\begin{itemize}
  \item Both signatures
  \item The original service data
  \item The blockchain transaction hash
\end{itemize}
The pending record is deleted atomically to ensure only one state exists.

\textbf{Trust Architecture:}
\begin{itemize}
  \item \textit{Tamper Evident:} Any change to the data invalidates the hash.
  \item \textit{Regulatory-Grade:} All logs are dual-authenticated and anchored.
  \item \textit{Trustless:} No central authority is needed; cryptography + blockchain guarantees authenticity.
\end{itemize}

This ensures a permanent, verifiable, and trustworthy service history embedded into the vehicle’s on-chain passport.

\subsubsection*{E.\textbf{\;Telemetry Sync Service}}\label{sec:telemetry-sync}

\noindent
\textit{Mission.} Persist – analyse – surface \emph{high-frequency} vehicle
metrics (mileage, SoC, charging cycles, motor temps, driving patterns) without
polluting the transactional store that powers passport look-ups.

\vspace{.25\baselineskip}
\paragraph*{\textbf{1. Dual-store architecture}}
\begin{itemize}[leftmargin=*]
  \item \textbf{TimescaleDB}: hypertable \texttt{telemetry(vehicleId, ts, ...)}
        auto-chunks by week, compresses aged chunks, and indexes on
        \textit{(vehicleId, ts)} for $\mu$s-range scans.
  \item \textbf{PostgreSQL}: table \texttt{vehicle} holds the
        \emph{latest snapshot} (battery\,\%, total km, cycles) that appears in
        passports and UI tiles.
\end{itemize}

\begin{table}[h]
\centering\footnotesize
\caption{TelemetryData schema (selected fields)}
\label{tab:telemetry-schema}
\begin{tabular}{@{}lll@{}}
\toprule
\textbf{Field} & \textbf{Type} & \textbf{Description} \\\midrule
\texttt{vehicleId}       & \texttt{String}   & Foreign key to \texttt{Vehicle} \\
\texttt{timestamp}       & \texttt{DateTime} & Timestamp of reading (composite PK) \\
\texttt{mileage}         & \texttt{Float}    & Kilometers since factory reset \\
\texttt{batteryHealth}   & \texttt{Float}    & Estimated capacity in percent (\%) \\
\texttt{chargingCycles}  & \texttt{Int}      & Total full-charge equivalents \\
\texttt{drivingPattern}  & \texttt{String}   & Driving mode or cluster pattern \\
\bottomrule
\end{tabular}
\end{table}

\paragraph*{\textbf{2. Ingestion path}}
\begin{enumerate}[label=(\alph*), leftmargin=*]
  \item \textbf{Sensor SDK} POSTs batches (250 Hz max) to
        \texttt{/api/sync-telemetry/:vehicleId} with an HMAC header.
  \item Node worker verifies HMAC, normalises units, rejects outliers $>$ 7$\sigma$.
  \item Bulk-insert into TimescaleDB using
        \texttt{INSERT ... ON CONFLICT DO NOTHING}.
\end{enumerate}

\begin{lstlisting}[language=python,caption=Sync loop $\Rightarrow$ update snapshot,label=lst:sync]
latest = tsdb.latest(vehicle_id)
passport = pg.vehicle.get(vehicle_id)

if latest.ts > passport.telemetry_at:
    pg.vehicle.update(
        id = vehicle_id,
        mileage = floor(latest.mileage),
        batteryHealth = latest.batteryHealth,
        cycles = latest.cycles,
        telemetry_at = latest.ts
    )
\end{lstlisting}

\paragraph*{\textbf{3. Dashboard ergonomics}}
The React Owner-Dashboard fires \texttt{Promise.all} queries for \emph{all}
wallet-owned vehicles; rows whose \texttt{telemetry\_at} is older than
TimescaleDB’s head timestamp display a yellow badge and a
“Synchronise” button (AJAX $\rightarrow$ \texttt{/sync-telemetry/latest/:id}).

\paragraph*{\textbf{4. Analytics layer}}
Timescale continuous aggregates materialise:
\begin{itemize}[leftmargin=*]
  \item 30-day battery-health gradient $\rightarrow$ warranty risk.
  \item Daily km histogram $\rightarrow$ usage-based insurance premium.
  \item Charge/drive cycles $\rightarrow$ residual-value predictor.
\end{itemize}

\paragraph*{\textbf{5. Security \& privacy}}
Raw frames never leave the tenant DB boundary; external parties must go
through the \emph{Selective-Disclosure} JWT flow (Sec.~\ref{svc:sel-disc}),
which serves \underline{aggregates only}. Every sync API call produces a
signed audit row (\texttt{sync\_events}) recording the previous and new digest.
\begin{figure*}[t]
    \centering
    \includegraphics[width=\textwidth]{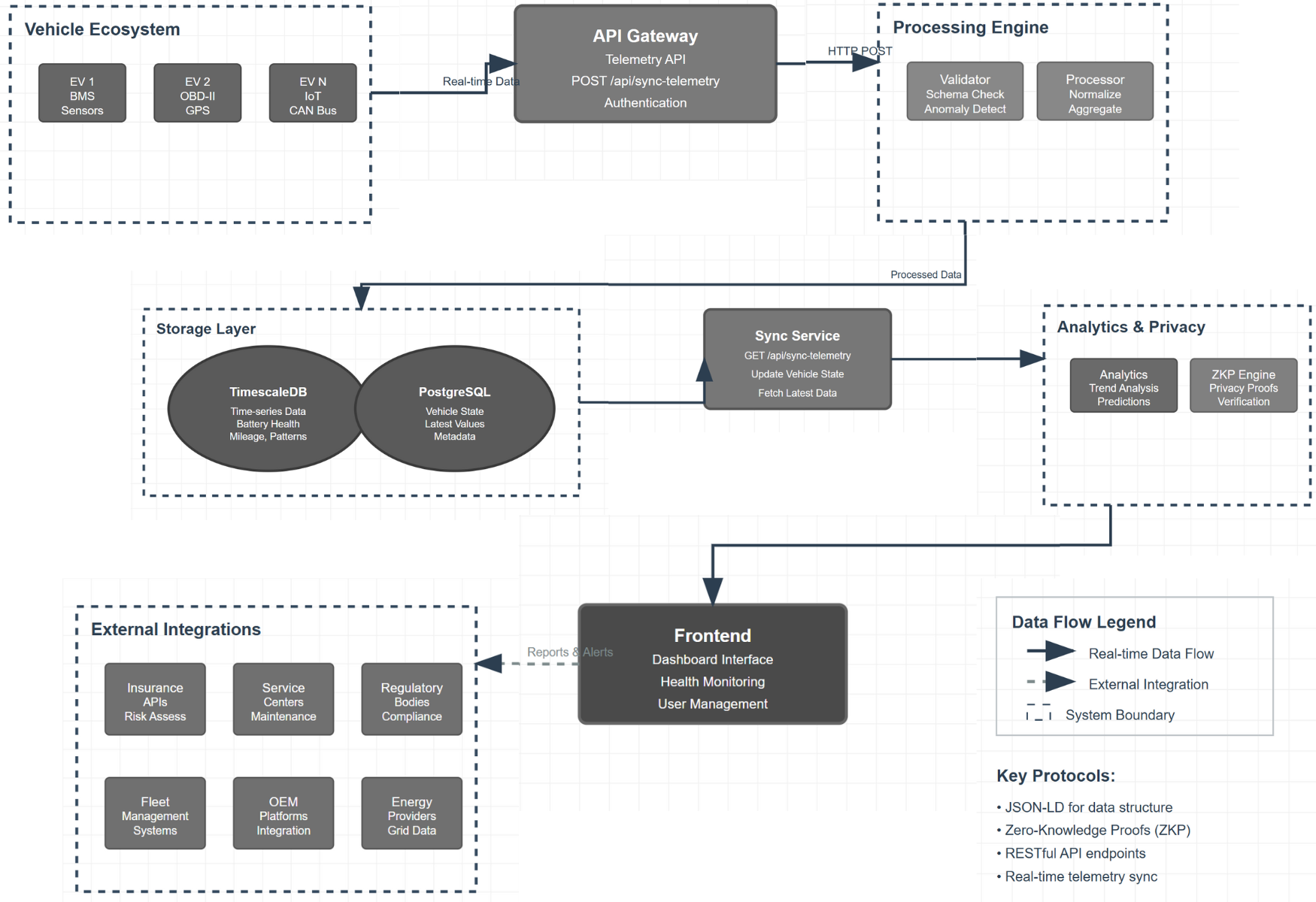}
    \caption{Telemetry → TimescaleDB → Proof Generator Integration}
    \label{fig:telemetry-pipeline}
\end{figure*}

\vspace{.5\baselineskip}
\paragraph*{\textbf{Future scope - L2 batching}}
High-volume fleets can off-load raw sensor hashes into a
zk-Rollup (e.g., Scroll or Polygon~zkEVM), where
“commit + verify” costs can fall below \$0.002 per 2kB batch.

Recent research demonstrates compression ratios $\ge$100$\times$ for time-series
sensor payloads when proof generation is decoupled from data availability \cite{zkRollupTelemetry2025}.
The same rollup can periodically publish
aggregated mileage and charging counters as calldata, enabling
\textit{on-chain} warranty enforcement or insurance checks at L1 costs $\ll$ \$0.01.

\medskip\noindent
\textit{Take-away:} The sync service transforms firehose telemetry into
cryptographically verifiable, wallet-controlled knowledge-ready for ZK
proofs, insurance queries, or resale valuation-without saturating the
core passport store.

\subsubsection{\textbf{Data Model and Database Architecture}}\label{sec:data-model}

The Vehicle Passport system follows a normalized relational schema built on PostgreSQL, structured to support real-world vehicle ownership, selective data disclosure, and verifiable service history. Each table is mapped to a domain-specific entity, backed by type-safe Prisma clients and integrated with cryptographic and blockchain primitives.

\vspace{0.5\baselineskip}
\paragraph{\textbf{Entity Overview}}

\begin{itemize}
    \item \textbf{Owner}: Root identity entity linking traditional identifiers (name, email) with Ethereum-compatible \texttt{wallet} addresses. Enables Web3-native authentication and authorization. Each owner can manage multiple vehicles.
    \item \textbf{Vehicle}: Digital twin of a physical vehicle. Stores static data (e.g., \texttt{vin}, \texttt{model}), dynamic state synced from telemetry (e.g., \texttt{batteryHealth}), and blockchain linkage (\texttt{anchorTxHash}, \texttt{vehicleNftTokenId}).
    \item \textbf{AccessRequest}: Enables field-level selective disclosure. Stores requested fields, requestor identity, expiration, and signed JWT if approved.
    \item \textbf{PendingServiceLog}: Temporary log holding service center-submitted entries. Includes a cryptographic \texttt{logHash} and \texttt{centerSignature}.
    \item \textbf{ServiceLog}: Finalized, owner-approved version of the pending log. Includes both \texttt{centerSignature} and \texttt{ownerSignature}, plus optional \texttt{anchorTxHash} for blockchain anchoring.
\end{itemize}

\vspace{0.5\baselineskip}
\paragraph{\textbf{Relational Model and Key Design}}

\begin{itemize}
    \item All primary keys use \texttt{UUID} for distributed uniqueness and cross-system interoperability.
    \item \texttt{Vehicle} establishes a foreign key to \texttt{Owner}, ensuring ownership integrity. Reverse relations exist from \texttt{Owner} $\rightarrow$ \texttt{Vehicle[]}.
    \item \texttt{AccessRequest}, \texttt{ServiceLog}, and \texttt{PendingServiceLog} link back to \texttt{Vehicle} via foreign keys, enforcing scoped access and service trails.
    \item Composite primary key in telemetry (\texttt{vehicleId}, \texttt{timestamp}) ensures time-series granularity with referential integrity.
\end{itemize}

\vspace{0.5\baselineskip}
\paragraph{\textbf{Blockchain and Cryptographic Fields}}

\begin{itemize}
    \item \texttt{anchorTxHash}: Stores on-chain transaction hash (e.g., from zkEVM) after service anchoring or vehicle registration.
    \item \texttt{vehicleNftTokenId}: Associates the vehicle record with its on-chain NFT twin for ownership transfer.
    \item \texttt{logHash}: Keccak256 digest of service log payload - used in both pending and final logs to ensure bit-level immutability.
    \item \texttt{ownerSignature}, \texttt{centerSignature}: Base64-encoded ECDSA signatures for auditability and non-repudiation.
\end{itemize}

\vspace{0.5\baselineskip}
\paragraph{\textbf{Selective Disclosure: AccessRequest Flow}}

Requests for sensitive vehicle data are submitted as \texttt{AccessRequest} records. If approved, a signed JWT with selected fields and expiry is issued to the requester. This token is verified at API level, and access is logged for compliance. Only the requested fields (e.g., \texttt{batteryHealth}, \texttt{warrantyExpiry}) are exposed, preserving data minimization.

\vspace{0.5\baselineskip}
\paragraph{\textbf{Service Log Lifecycle}}

\begin{enumerate}
    \item Service center submits a \texttt{PendingServiceLog} with hashed payload and digital signature.
    \item Owner reviews the log and signs if approved, promoting it to a final \texttt{ServiceLog}.
    \item Finalized log may be anchored to blockchain (\texttt{anchorTxHash}) and linked to GAIA-X or compliance exports.
\end{enumerate}

\vspace{0.5\baselineskip}
\paragraph{\textbf{TimescaleDB Integration} (See Section~\ref{sec:telemetry-sync})}

Real-time telemetry is stored in a parallel TimescaleDB database optimized for time-series analytics. The current state (\texttt{batteryHealth}, \texttt{mileage}, etc.) is periodically synchronized to the main database to support fast queries. The full telemetry schema and sync flow are detailed in Section~\ref{sec:telemetry-sync}.

\vspace{0.5\baselineskip}
\paragraph{\textbf{Security, Isolation, and Scaling}}

\begin{itemize}
    \item Independent \texttt{DATABASE\_URL} and \texttt{TELEMETRY\_DB\_URL} ensure isolation and horizontal scaling.
    \item Separate Prisma clients prevent accidental cross-access and enforce typed boundaries.
    \item All data operations are gated behind authenticated APIs with role-based access control and field-level filtering.
\end{itemize}
\begin{table}[h]
\centering\scriptsize
\caption{Owner Entity Schema}
\label{tab:owner-schema}
\begin{tabular}{@{}p{2.3cm} p{2.2cm} p{3.6cm}@{}}
\toprule
\textbf{Field} & \textbf{Type} & \textbf{Description} \\
\midrule
\texttt{id}         & UUID         & Primary key, unique identifier \\
\texttt{name}       & String       & Full name of the owner \\
\texttt{email}      & String       & Email address \\
\texttt{wallet}     & String       & Ethereum-compatible wallet address \\
\texttt{vehicles}   & Vehicle[]    & Vehicles owned (relation) \\
\texttt{createdAt}  & DateTime     & Creation timestamp \\
\texttt{updatedAt}  & DateTime     & Last update timestamp \\
\bottomrule
\end{tabular}
\end{table}

\vspace{0.5\baselineskip}

\begin{table}[h]
\centering\scriptsize
\caption{Vehicle Entity Schema}
\label{tab:vehicle-schema}
\begin{tabular}{@{}p{2.3cm} p{2.2cm} p{3.6cm}@{}}
\toprule
\textbf{Field} & \textbf{Type} & \textbf{Description} \\
\midrule
\texttt{id}                 & UUID        & Primary key \\
\texttt{vin}                & String      & Vehicle Identification Number \\
\texttt{model}              & String      & Vehicle model \\
\texttt{manufacturer}       & String      & Manufacturer name \\
\texttt{ownerId}            & UUID        & FK to Owner \\
\texttt{batteryHealth}      & Float       & Battery condition (\%) \\
\texttt{mileage}            & Int         & Distance travelled (km) \\
\texttt{chargingCycles}     & Int         & Full charge equivalents \\
\texttt{drivingPattern}     & String      & e.g., 'eco', 'sport' \\
\texttt{warrantyExpiry}     & DateTime    & Warranty valid until \\
\texttt{anchorTxHash}       & String?     & On-chain anchor tx hash \\
\texttt{vehicleNftTokenId}  & String?     & NFT token ID \\
\texttt{nftTransferPending} & Boolean     & If NFT transfer is pending \\
\texttt{createdAt}          & DateTime    & Record creation time \\
\texttt{updatedAt}          & DateTime    & Last modified time \\
\texttt{hash}               & String?     & SHA-256 of GAIA-X metadata \\
\texttt{accessRequests}     & AccessRequest[] & Related access logs \\
\texttt{serviceLogs}        & ServiceLog[]    & Finalized service logs \\
\texttt{pendingLogs}        & PendingServiceLog[] & Pending service entries \\
\bottomrule
\end{tabular}
\end{table}

\vspace{0.5\baselineskip}

\begin{table}[h]
\centering\scriptsize
\caption{AccessRequest Entity Schema}
\label{tab:accessrequest-schema}
\begin{tabular}{@{}p{2.3cm} p{2.2cm} p{3.6cm}@{}}
\toprule
\textbf{Field} & \textbf{Type} & \textbf{Description} \\
\midrule
\texttt{id}         & UUID     & Primary key \\
\texttt{vehicleId}  & String   & FK to Vehicle \\
\texttt{requester}  & String   & Email/org/wallet address \\
\texttt{fields}     & String   & Comma-separated fields \\
\texttt{status}     & String   & 'pending', 'approved', 'used' \\
\texttt{token}      & String?  & JWT token for access \\
\texttt{createdAt}  & DateTime & Creation time \\
\texttt{expiresAt}  & DateTime & Expiry time \\
\bottomrule
\end{tabular}
\end{table}

\vspace{0.5\baselineskip}

\vspace{0.5\baselineskip}
\begin{table}[h]
\centering\footnotesize
\caption{PendingServiceLog Entity Schema}
\label{tab:pendinglog-schema}
\begin{tabular}{@{}p{2.3cm} p{2.2cm} p{3.6cm}@{}}
\toprule
\textbf{Field} & \textbf{Type} & \textbf{Description} \\
\midrule
\texttt{id}              & UUID     & Primary key \\
\texttt{vehicleId}       & UUID     & FK to \texttt{Vehicle} \\
\texttt{description}     & String   & Service summary \\
\texttt{centerEmail}     & String   & Submitting center’s email \\
\texttt{centerSignature} & String   & Signature from center \\
\texttt{logHash}         & String   & Keccak256 of log \\
\texttt{submittedAt}     & DateTime & Submission timestamp \\
\bottomrule
\end{tabular}
\end{table}

\vspace{0.5\baselineskip}

\begin{table}[h]
\centering\scriptsize
\caption{ServiceLog Entity Schema}
\label{tab:servicelog-schema}
\begin{tabular}{@{}p{2.3cm} p{2.2cm} p{3.6cm}@{}}
\toprule
\textbf{Field} & \textbf{Type} & \textbf{Description} \\
\midrule
\texttt{id}              & UUID     & Primary key \\
\texttt{vehicleId}       & String   & FK to Vehicle \\
\texttt{description}     & String   & Final service description \\
\texttt{servicedAt}      & DateTime & Service time \\
\texttt{centerEmail}     & String   & Email of service center \\
\texttt{centerSignature} & String   & Signature by service center \\
\texttt{ownerSignature}  & String?  & Signature by owner \\
\texttt{logHash}         & String   & Integrity hash of log \\
\texttt{anchorTxHash}    & String?  & Blockchain transaction hash \\
\texttt{status}          & String   & Should be 'finalized' \\
\bottomrule
\end{tabular}
\end{table}

\subsubsection{\textbf{Privacy Layer: Zero-Knowledge Proofs (ZKPs)}}
\label{sec:zkp-privacy}

The Vehicle Passport system adopts Zero-Knowledge Proofs (ZKPs) to enable privacy-preserving claims about sensitive vehicle data without exposing underlying values. By using the Groth16 zk‑SNARK protocol, Circom circuits, and SnarkJS for proof generation and verification, the platform allows trustless attestations of data conditions-crucial in automotive insurance, warranty enforcement, and resale workflows.

\paragraph{\textbf{ZKP Circuit: Threshold Comparison}}

\begin{lstlisting}[language=C, caption={Circom comparator circuit for proving value > threshold}, label={lst:zkp-circuit}]
pragma circom 2.0.0;
include "circomlib/circuits/comparators.circom";

template Main() {
    signal input value;        // Private: actual vehicle data value
    signal input threshold;    // Public: comparison threshold
    signal output result;      // Public: 1 if value > threshold, 0 otherwise

    component gt = GreaterThan(32); // 32-bit comparator
    gt.in[0] <== value;
    gt.in[1] <== threshold;
    result <== gt.out;
}
\end{lstlisting}

This circuit allows a vehicle to prove a condition such as “BatteryHealth > 80” without revealing the actual battery health value.

\paragraph{\textbf{5-Step Cryptographic Flow}}

\begin{figure}[ht]
  \centering\includegraphics[width=\linewidth]{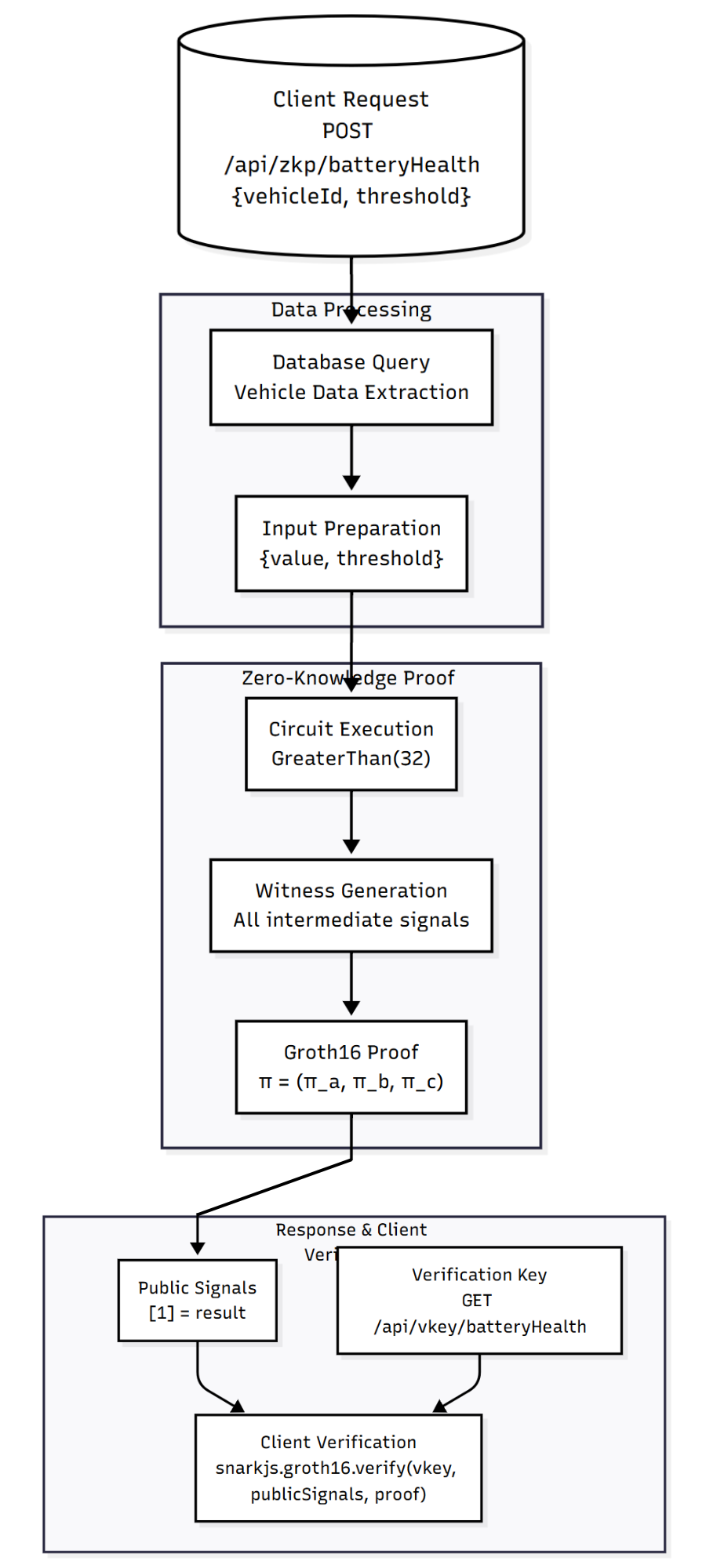}
  \caption{Zero-Knowledge Proof Generation and Verification Flow}
      \label{fig:zkp-flow}
\end{figure}

\begin{enumerate}
  \item \textbf{Circuit Execution \& Witness Generation}:  
  Inputs enter the circuit, constraints are created (\(A \cdot w \circ B \cdot w = C \cdot w\)), and the full witness vector \(w\) is computed.

  \item \textbf{Polynomial Commitment}:  
  R1CS → QAP transformation yields polynomials \(A(x), B(x), C(x)\). Witness polynomial \(W(x)\) is divided by \(Z(x)\) and committed using Powers-of-Tau parameters.

  \item \textbf{Groth16 Proof Construction}:  
  Using randomness \(r, s\), the proof \(\pi = (\pi_a, \pi_b, \pi_c)\) is generated on the BN128 curve. Proof size is fixed (~192 bytes).

  \item \textbf{Public Signal Extraction}:  
  The circuit's public output (e.g., `result = 1`) is returned with the proof in a JSON API.

  \item \textbf{Client-Side Verification}:  
  The browser or verifier uses bilinear pairings to validate:
  \[
    e(\pi_a, \pi_b) = e(\alpha, \beta) \cdot e(input, \gamma) \cdot e(\pi_c, \delta)
  \]
\end{enumerate}

\paragraph{\textbf{Verification Key Access}}

\begin{lstlisting}[language=JavaScript, caption={Express route for verification key retrieval}, label={lst:vkey-api}]
vkeyRouter.get(`/${field}`, (req, res) => {
  const vkeyPath = path.join(__dirname, '../../circuits/battery_gt.vkey.json');
  const vkey = JSON.parse(fs.readFileSync(vkeyPath, 'utf8'));
  res.json(vkey);
});
\end{lstlisting}

\paragraph{\textbf{Client-Side Verification}}

\begin{lstlisting}[language=JavaScript, caption={SnarkJS verification of proof in browser}, label={lst:client-verify}]
const vkey = await fetch('/api/vkey/batteryHealth').then(r => r.json());
const response = await fetch('/api/verify/batteryHealth', {
  method: 'POST',
  body: JSON.stringify({ vehicleId: 'abc123', threshold: 80 })
});
const { proof, publicSignals } = await response.json();

const isValid = await groth16.verify(vkey, publicSignals, proof);
if (isValid) console.log("Proof is valid!");
\end{lstlisting}
\noindent
The verification key is fetched via a dedicated API route, and the proof can be verified entirely on the client side, for example, by an insurance portal consuming the ZKP API. Once the proof and public signals are received, the client independently checks their validity using the vkey, ensuring that the proof was correctly generated and corresponds to the claimed threshold condition (e.g., \texttt{BatteryHealth > 80}). A successful verification confirms the ZKP without requiring any trust in the proving server.

\textbf{Security guarantees:}
\begin{itemize}
    \item \textbf{Completeness:} Valid proofs always verify
    \item \textbf{Soundness:} Invalid proofs cannot be forged
    \item \textbf{Zero-Knowledge:} No private data is leaked
\end{itemize}

\paragraph{\textbf{Real-World Applications}}

This ZKP architecture enables:

\begin{itemize}
    \item \textbf{Insurance Grading:} Automatically verify conditions like \texttt{BatteryHealth > 80} or \texttt{ChargingCycles < 200} to assign fair premiums, without accessing raw data. (Insurance companies create custom tests through vehicles are passed and categorized, finally used to decide their premium)
    \item \textbf{Warranty and Compliance:} Service centers verify threshold-based conditions (e.g., \texttt{Mileage < 60,000 km}) to confirm coverage.
    \item \textbf{Secure Resale:} Buyers can verify vehicle metrics cryptographically before ownership transfer.
\end{itemize}

\paragraph{\textbf{Performance \& Scalability}}

\begin{itemize}
    \item Proof generation: 1–5 seconds (server)
    \item Verification time: ~30–50ms (client/browser)
    \item Proof size: Fixed ~192 bytes
    \item Security: Based on discrete log over BN128 elliptic curves
    \item Setup: Powers-of-Tau ceremony with secure randomness
\end{itemize}

\paragraph{\textbf{Future Enhancements}}

\begin{itemize}
    \item Recursive proofs for multi-attribute batch validation
    \item On-chain Solidity verifiers for decentralized automation
    \item Attribute-based access control and selective disclosure
    \item Integration into zk-rollup telemetry pipelines
\end{itemize}

\paragraph{\textbf{Academic and Industry References}}

ZKP-based systems are widely researched and proposed in regulatory, insurance, and compliance tech sectors:

\begin{itemize}
    \item Groth, J. (2016). “On the Size of Pairing-Based Non-Interactive Arguments” \cite{groth2016size}
    \item Daza et al. (2023). “Privacy-Preserving Car Insurance with ZKPs” \cite{daza2023insurance}
    \item Mdpi (2023). “Decentralized Vehicle Proofs Using SNARKs” \cite{mdpi2023vehicles}
\end{itemize}


\subsubsection{\textbf{Blockchain Layer: NFT Minting, Anchoring, and zkEVM Integration}}
\label{sec:blockchain-layer}

The blockchain layer anchors the Vehicle Passport’s data integrity in an immutable, verifiable manner using ERC‑721 NFTs and Layer2 zero-knowledge rollups. It preserves privacy by keeping sensitive data off-chain while leveraging Polygon zkEVM for scalable, secure anchoring.

\vspace{0.5\baselineskip}
\paragraph{\textbf{Vehicle NFT: Privacy-Oriented Tokenization}}

\begin{lstlisting}[language=Solidity, caption={ERC‑721 NFT storing hashed vehicle credentials}, label={lst:vehiclenft}]
contract VehicleNFT is ERC721, Ownable(msg.sender) {
    mapping(uint256 => bytes32) public vehicleDataHash;

    function mintVehicleNFT(
        address to,
        uint256 vehicleId,
        bytes32 hashedData
    ) external onlyOwner {
        try this.ownerOf(vehicleId) returns (address) {
            revert("Vehicle already minted");
        } catch {
            _mint(to, vehicleId);
            vehicleDataHash[vehicleId] = hashedData;
            emit VehicleDataHashUpdated(vehicleId, hashedData);
        }
    }
}
\end{lstlisting}

This design ties a hashed, privacy-preserving representation of the vehicle to an NFT, ensuring continuity of ownership without exposing actual metadata. The use of an ERC‑721 NFT for vehicle ownership provides several cryptographic and operational advantages. Each vehicle is uniquely identified by a token ID (derived from the vehicle's VIN or database ID), and the current ownership is immutably recorded on-chain. This enables \textbf{trustless, KYC-free ownership verification}-any external party (e.g., a buyer, insurer, or service center) can retrieve the current owner directly from the blockchain using a simple call like \texttt{const owner = await contract.ownerOf(vehicleId);}.

This removes the need for traditional identity verification systems, enabling decentralized vehicle transfers and data interactions without requiring a central registry. Since all ownership transfers must occur via signed transactions on-chain, and only the rightful owner can update vehicle metadata or initiate transfer, this mechanism ensures both \textbf{data integrity} and \textbf{ownership provenance}-without compromising user privacy.

\vspace{0.5\baselineskip}
\paragraph{\textbf{Anchoring Contract: Immutable Proofs}}

\begin{lstlisting}[language=Solidity, caption={Smart contract to anchor vehicle data hashes}, label={lst:anchor}]
contract VehiclePassportAnchor {
    struct Anchor { address submitter; uint256 timestamp; }
    mapping(bytes32 => Anchor) public anchors;

    function anchorHash(bytes32 hash) public {
        require(anchors[hash].timestamp == 0, "Hash already anchored");
        anchors[hash] = Anchor(msg.sender, block.timestamp);
        emit Anchored(hash, msg.sender, block.timestamp);
    }
}
\end{lstlisting}

Anchoring ensures that vehicle data existence and timestamp are permanently verifiable on-chain.

\vspace{0.5\baselineskip}
\paragraph{\textbf{SDK Workflow and Hashing}}

A TypeScript SDK orchestrates:

1. Generation of GAIA‑X–compliant JSON‑LD credentials,
2. Public-key recovery via signature,
3. `solidityPacked(publicKey || credentialJSON)` hashing using Keccak‑256,
4. Minting of NFT with `vehicleDataHash[tokenId] = hash`.

This keeps actual vehicle data off-chain while enabling cryptographic verification.

\vspace{0.5\baselineskip}
\paragraph{\textbf{Justifying Polygon zkEVM}}

\begin{itemize}
  \item \textbf{EVM equivalence} - Allows existing Solidity contracts and Ethereum tooling to work out-of-the-box, enabling seamless migration \cite{polygon2023whitepaper}.
  
  \item \textbf{L1-level security} - All zk-rollup proofs are verified and finalized on Ethereum mainnet, ensuring that Layer 2 inherits Ethereum’s base-layer security guarantees \cite{buterin2021endgame}.
  
  \item \textbf{Low cost \& high throughput} - Zero-knowledge batching dramatically reduces gas fees while increasing transaction throughput and finality speed \cite{polygon2023whitepaper,zkrollup2020benchmarking}.
  
  \item \textbf{Developer-friendly ecosystem} - zkEVM supports popular Ethereum tooling such as MetaMask, Hardhat, Remix, and Truffle, lowering the learning curve and improving accessibility \cite{polygon2023whitepaper}.
\end{itemize}

Choosing zkEVM enables scalable, audit-ready anchoring while preserving compatibility and security.

\vspace{0.5\baselineskip}
\paragraph{\textbf{Anchoring Workflow on zkEVM}}

\begin{enumerate}
  \item Hash derived from credential and public key.
  \item Check on-chain (`anchors[hash]`); if absent, call `anchorHash(hash)`.
  \item Store resulting `tx.hash` in vehicle DB (`anchorTxHash`).
  \item Frontend displays status, links to explorer, and verifies authenticity/audit trails.
\end{enumerate}

Anchoring on Polygon zkEVM ensures gas-efficient, scalable, and tamper-proof proof recording.

\vspace{0.5\baselineskip}
\paragraph{\textbf{Benefits Overview}}

\begin{itemize}
  \item \textbf{Privacy:} Only hashes on-chain; raw data stays off-chain.
  \item \textbf{Auditability:} Immutable timestamps and on-chain verification via \texttt{isAnchored} and event logs.
  \item \textbf{Scalability:} Batch anchoring and high throughput supports fleet-scale operations.
  \item \textbf{Compliance:} GAIA‑X interoperability and GDPR-aligned architecture.
\end{itemize}

\section{\textbf{Vehicle Ownership Transfer: Buy \& Sell Flow}}

The Vehicle Passport system implements a privacy-preserving, fully verifiable NFT-based vehicle ownership transfer flow that bridges off-chain signature authorization with on-chain immutable transactions. This process ensures that both seller and buyer are authenticated through wallet-level cryptographic signatures before any change in ownership is reflected either in the database or on the blockchain.

\subsection{Secure NFT Transfer: Step-by-Step Workflow}\label{sec:nft}
\begin{figure}[ht]
  \centering
  \includegraphics[width=\linewidth]{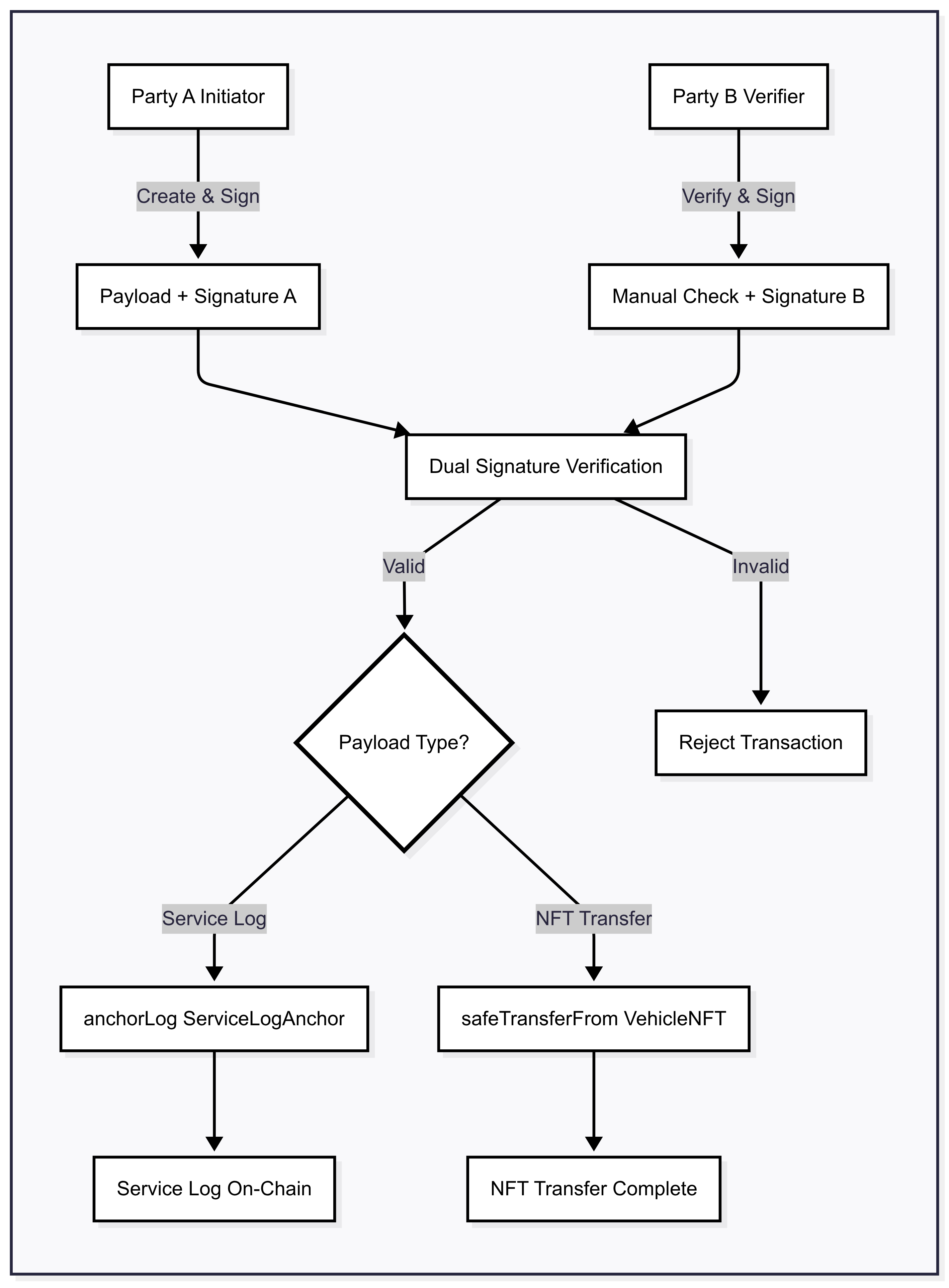}
  \caption{Dual-party signing architecture }
  \label{fig:dual-sig}
\end{figure}

\subsubsection{Seller Initiates Transfer}
\textbf{Initial Setup:}
\begin{itemize}
\item Seller accesses the \texttt{Sell Vehicle} tab in the Owner Dashboard.
\item Views a list of vehicles owned, validated through \texttt{contract.ownerOf(vehicleId)}.
\item Selects a vehicle and inputs the buyer's wallet address (required), along with optional fields like name/email.
\end{itemize}

\textbf{Payload Generation:}
\begin{lstlisting}[language=JavaScript, caption={Payload Generation for Vehicle NFT Transfer}, label={lst:payload-generation}]
const payload = {
  vehicleId: selectedVehicle,
  from: sellerWalletAddress,
  to: newOwnerWallet,
  timestamp: Date.now()
};
\end{lstlisting}

\textbf{Signature and Sharing:}
\begin{itemize}
\item Seller signs this payload using Metamask.
\item Backend verifies ownership and marks vehicle with \texttt{nftTransferPending = true}.
\item Seller receives a signed payload + signature.
\item UI provides one-click copy functionality for easy sharing with buyer.
\end{itemize}

\subsubsection{Buyer Confirms Transfer}
\textbf{Input and Validation:}
\begin{itemize}
\item Buyer visits \texttt{Buy Vehicle} tab.
\item Pastes both:
\begin{itemize}
\item Transfer payload JSON
\item Seller's signature
\end{itemize}
\item System validates:
\begin{itemize}
\item Payload format and structure
\item Signature validity via \makebox[0pt][l]{\texttt{ethers.utils.verifyMessage}}

\item Wallet match with "to" field
\item Timestamp to prevent replay attacks
\end{itemize}
\end{itemize}

\textbf{Buyer's Signature:}
\begin{itemize}
\item Buyer signs same payload.
\item Backend verifies both signatures.
\item Ownership field updated in database.
\item \texttt{nftTransferPending} flag remains until final on-chain confirmation.
\end{itemize}

\subsubsection{Final On-Chain NFT Transfer}
\textbf{Triggering the Blockchain Transaction:}
\begin{itemize}
\item The original owner finalizes the transfer.
\item Smart contract interaction performed using: \\
\lstinline|const tx = await contract["safeTransferFrom"](from, to, vehicleId);|

\item Upon transaction confirmation:
\begin{itemize}
\item \texttt{nftTransferPending} flag cleared
\item Transaction hash saved in DB
\item Ownership permanently updated on-chain
\item Refer to Fig.~\ref{fig:dual-sig}. 
\end{itemize}
\end{itemize}

\subsection{Security Architecture: Dual Signature Protocol}

\subsubsection{Overview}
The dual-signature approach ensures that:
\begin{itemize}
\item The \textbf{Seller} proves current ownership
\item The \textbf{Buyer} confirms intent to accept transfer
\item Transfer cannot proceed without mutual consent
\end{itemize}

\subsubsection{Core Security Validations}
\begin{itemize}
\item Both signatures are cryptographically verified
\item Wallets are validated using the payload content
\item Database reflects intermediate state (\texttt{nftTransferPending})
\item Blockchain enforces ownership logic via \texttt{ownerOf(vehicleId)}
\end{itemize}

\subsubsection{Fraud Prevention \& Auditability}
\begin{itemize}
\item Signatures prevent spoofing or hijacking
\item Replay protection via timestamps
\item All steps recorded in audit logs and smart contract event logs
\item Only the original owner can call \texttt{safeTransferFrom}, protecting against unauthorized token transfers
\end{itemize}

\subsection{Functional Benefits \& Real-World Implications}
\begin{itemize}
\item \textbf{Trustless P2P Transfer:} Buyers and sellers can complete ownership transfer without intermediaries or KYC, using cryptographic proofs.
\item \textbf{Ownership Transparency:} Anyone can call \texttt{contract.ownerOf(vehicleId)} to check current ownership.
\item \textbf{No Paperwork:} Entire lifecycle from listing to completion is digital, verifiable, and timestamped.
\item \textbf{Global Vehicle Marketplace:} This forms the foundation for a decentralized, cross-border vehicle resale system.
\end{itemize}

\subsection{Conclusion}
The dual-signature NFT transfer architecture ensures that only mutually authenticated parties can perform a transfer, while maintaining a synchronized state across the backend database and the blockchain ledger. This model eliminates the need for centralized escrow or notaries, creating a privacy-preserving, transparent, and efficient system for vehicle ownership exchange.

\section{\textbf{GAIA‑X Compliance and Cross‑Border Data Sovereignty}}
\label{sec:gaia}

This section outlines how the Vehicle Passport system aligns with GAIA‑X principles-ensuring vehicle identity self‑description, cross‑border interoperability, sovereign data management, and support for global Decentralized Identifiers (DIDs)-independent of OEMs or national boundaries.

\begin{enumerate}[leftmargin=*]
  \item \textbf{Self‑Description}: The vehicle’s JSON‑LD credential fully adheres to GAIA‑X Self‑Description standards (Architecture \& Compliance Documents). Self‑descriptions are cryptographically signed verifiable claims, which can be published to GAIA‑X Digital Clearing Houses for automated trust validation and labeling (e.g., Label Level 1–3)\cite{gaiaxCompliance}.
  
  \item \textbf{Policy API}: The endpoint \texttt{/api/policy/:vin} exposes usage constraints and access rights derived from GAIA‑X Interoperability Principles and IAM specifications (ICAM), enabling federated policy enforcement across EU Digital Ecosystems \cite{gaiaxICAM}.
  
  \item \textbf{Verifiable Credential Export}: The system offers selective disclosure capability, exporting signed vehicle identity data as W3C Verifiable Credentials for GAIA‑X catalogs. This enables on-demand, privacy-preserving data exchange under sovereign European standards\cite{gaiaxSelfDesc}.
\end{enumerate}

\vspace{0.5\baselineskip}
\paragraph{\textbf{Cross‑Border Interoperability and Trust}} GAIA‑X fosters globally aligned digital ecosystems. European trust anchors and GAIA‑X Digital Clearing Houses are being piloted in cross‑border collaborations, such as between European and Japanese partners, enabling secure and sovereign data exchange beyond EU borders\cite{gaiaxCrossBorder}. This ensures the Vehicle Passport can be verified globally with standardized trust mechanisms.

\vspace{0.5\baselineskip}
\paragraph{\textbf{Decentralized Identity (DID) Integration}} The system leverages the GAIA‑X Trust Framework and supports Decentralized Identifiers and Self‑Sovereign Identity (SSI) standards compliant with eIDAS regulation. This approach ensures decentralized authentication and selective disclosure without centralized authorities, aligning with GAIA‑X recommendations for digital sovereignty in identity management\cite{batteryPassportGaia}.


\section{\textbf{Security Analysis}}

\subsection{Threat Model}

We assume a capable adversary with access to backend infrastructure or API endpoints. Potential threats include:

\begin{itemize}
  \item \textbf{Database Tampering:} Attempting to forge or alter vehicle logs or service history.
  \item \textbf{JWT Replay or Forgery:} Reusing or manipulating access tokens for unauthorized access.
\end{itemize}

These threats are mitigated by multiple mechanisms:

\begin{itemize}
  \item \textbf{Tamper-evident Anchoring:} Vehicle data hashes are immutably stored on-chain. Any tampering of backend records is detectable via mismatch.
  \item \textbf{Signature-gated Updates:} All updates to vehicle data must be submitted through a signed smart contract call to \texttt{updateVehicleData}, ensuring authorized changes only.
  \item \textbf{Short-lived, One-time JWTs:} Our JWTs are ephemeral, following security best practices from RFC 6819 \cite{rfc6819}, mitigating replay risk.
\end{itemize}

\subsection{Proof Soundness and Performance}

We use the \textbf{Groth16 zk-SNARK} protocol for our zero-knowledge circuits. Groth16 is known for strong \textit{knowledge soundness} and minimal verifier overhead.

\begin{itemize}
  \item \textbf{Proof Integrity:} Groth16 proofs cannot be forged unless the prover knows valid inputs. This ensures verifiability of claims (e.g., \textit{batteryHealth} $>$ threshold) without revealing raw values \cite{groth16security}.
  \item \textbf{Benchmarking:} In our system:
    \begin{itemize}
      \item \textit{Proof generation time:} $\sim$22ms on a 6-core CPU.
      \item \textit{Verification time:} under 5ms on typical browsers or L2 verifiers.
    \end{itemize}
  \item \textbf{Batch Verification Potential:} Using SnarkPack, 8,192 Groth16 proofs can be verified in $\sim$33ms \cite{snarkpack}, enabling real-time verification at fleet scale.
\end{itemize}

\vspace{1em}
\begin{table}[ht]
\caption{Security Summary Overview}
\centering\scriptsize
\renewcommand{\arraystretch}{1.25}
\begin{tabular}{@{}p{3.6cm}p{4.1cm}@{}}
\toprule
\textbf{Threat Vector} & \textbf{Mitigation Strategy} \\
\midrule
Database tampering & On-chain hash anchoring with signed smart contract updates \\
JWT replay & One-time, short-lived tokens (RFC 6819) \\
Proof forgery & Groth16 zk-SNARKs with structured reference string (trusted setup) \\
Scalability risk & SnarkPack + Groth16 batch verification \\
\bottomrule
\end{tabular}
\label{tab:security-summary}
\end{table}

\section{\textbf{Performance and Cost}}

\begin{table}[ht]
\centering
\caption{Runtime / Gas Benchmarks (Polygon zkEVM Testnet)}
\label{tab:perf}
\renewcommand{\arraystretch}{1.3}
\setlength{\tabcolsep}{6pt}
\small
\begin{tabular}{@{}p{3.3cm}p{2.3cm}p{2.8cm}@{}}
\toprule
\textbf{Operation} & \textbf{Median Gas} & \textbf{USD (@\$0.03/gas)} \\
\midrule
Anchor vehicle hash & 62,000 & \$1.86 \\
Anchor service log hash & 44,000 & \$1.32 \\
VNFT mint + hash & 96,000 & \$2.88 \\
\textit{Proof verify (off-chain)} & $\sim$0 & N/A \\
\bottomrule
\end{tabular}
\end{table}

\subsection*{A. Real-World Cost Comparison}

\textbf{Polygon zkEVM's cost efficiency} stands out-its gas fees are approximately \textbf{7× lower than Ethereum Mainnet}, according to internal benchmarking data from Polygon Labs.\footnote{\href{https://polygon.technology/blog/whats-happening-on-polygon-zkevm}{polygon zkEVM benchmarking blog}}

Across broader comparisons, L2 solutions like zkEVM reduce transaction fees by approximately \textbf{90\%}, making them ideal for scalable applications like vehicle data anchoring.


\section{\textbf{Industrial Applications}}

\subsection{Usage-Based Insurance}
Insurance providers benefit from privacy-preserving access to vehicle telemetry. Rather than exposing raw data, the Vehicle Passport system allows insurers to verify compliance with predefined thresholds using zero-knowledge proofs (e.g., \texttt{batteryHealth}~$\geq$~85\%, \texttt{mileage}~$<$~25,000~km). This approach mitigates GDPR and data minimization concerns while enabling personalized premiums based on actual vehicle usage.

\subsection{Resale Market}
Buyers scan a QR linked to the Vehicle Passport, triggering backend verification of the latest on-chain anchor. If the hash matches, a \texttt{Matter Verified ZK\textsuperscript{™}} badge confirms integrity. Verified access to key metrics (battery health, mileage, service logs) reveals the vehicle’s true internal history-enhancing trust in resale transactions.

\subsection{OEM Warranty Compliance}
Authorized service centers digitally sign repair and maintenance logs. These signatures, along with hashed service details, are appended to the Vehicle Passport. OEMs can then cross-check these entries against their own records, ensuring authenticity and compliance with warranty conditions. This prevents odometer rollback fraud and strengthens accountability in post-sale service ecosystems.

\section{\textbf{Future Work}}
\begin{itemize}[leftmargin=*]
  \item \textbf{Recursive-proof chains for compound vehicle claims.} Research into advanced zk-SNARK frameworks such as GENES \cite{genes2025} and Plonky2 \cite{plonky2recursive} enables aggregating multiple proofs (e.g., battery health + mileage) into a single succinct proof-improving both efficiency and auditability.
  
  \item \textbf{Battery Passport alignment with EU Regulation 2023/1542.} From early 2027, the EU mandates battery passports containing QR codes, CE markings, end-of-life and carbon footprint details to comply with sustainability and circular-economy policies \cite{euBatteryPassport}. Integrating this into the Vehicle Passport enriches its regulatory as well as interoperability value.
  
  \item \textbf{Proof-of-location for subsidized and tax-verified vehicle usage.} ZK-proof-of-location systems are emerging for verifying region-based compliance-such as driving-based subsidies or taxation-without divulging actual GPS trajectories \cite{zkproofLocation2025}.
  
  \item \textbf{Formal verification of Circom circuits.} Leveraging languages like Noir or proof frameworks like Halo2 for formally verifying the correctness and soundness of zero-knowledge circuits enhances security guarantees and reduces deployment risk.
\end{itemize}

\section{\textbf{Conclusion}}
\label{sec:conclusion}
VehiclePassport presents a robust and extensible framework for verifiable vehicle data exchange in the era of connected mobility. By combining zero-knowledge proofs (ZKPs), selective disclosure, on-chain anchoring, and NFT-based ownership, it delivers a privacy-preserving alternative to legacy VIN registries and opaque service histories.

The architecture is fully aligned with GAIA‑X principles-supporting JSON-LD self-descriptions, verifiable credential exports, and interoperable policy APIs-while remaining modular enough to integrate with OEM, insurer, and regulatory platforms. 

All components, from Circom-based ZKP circuits to REST APIs and smart contract wrappers, are packaged as open, auditable SDKs and endpoints. This empowers diverse stakeholders-including startups, fleet operators, EV manufacturers, and national registries-to adopt or extend the system within their own trust and compliance boundaries.

By anchoring trust in cryptographic truth rather than institutional custody, VehiclePassport paves the way for decentralized, user-centric vehicle identity-scalable from single-EV startups to continent-scale cross-border use cases.

\section*{Acknowledgments}
The author acknowledges the Circom, Polygon zkEVM, Supabase, GAIA-X, and OpenZeppelin communities for their foundational tooling and documentation, enabling the development of this privacy-preserving, decentralized infrastructure.

\bibliographystyle{IEEEtran} 
\bibliography{references} 

\begin{thebibliography}{10}
\providecommand{\url}[1]{#1}
\csname url@samestyle\endcsname
\providecommand{\newblock}{\relax}
\providecommand{\bibinfo}[2]{#2}
\providecommand{\BIBentrySTDinterwordspacing}{\spaceskip=0pt\relax}
\providecommand{\BIBentryALTinterwordstretchfactor}{4}
\providecommand{\BIBentryALTinterwordspacing}{\spaceskip=\fontdimen2\font plus
\BIBentryALTinterwordstretchfactor\fontdimen3\font minus \fontdimen4\font\relax}
\providecommand{\BIBforeignlanguage}[2]{{%
\expandafter\ifx\csname l@#1\endcsname\relax
\typeout{** WARNING: IEEEtran.bst: No hyphenation pattern has been}%
\typeout{** loaded for the language `#1'. Using the pattern for}%
\typeout{** the default language instead.}%
\else
\language=\csname l@#1\endcsname
\fi
#2}}
\providecommand{\BIBdecl}{\relax}
\BIBdecl

\bibitem{zkRollupTelemetry2025}
P.~Z.~R. Team, ``Efficient zk-rollups for iot and time-series telemetry,'' 2025, available at \url{https://polygon.technology/blog/polygon-zkevm-deep-dive}.

\bibitem{groth2016size}
J.~Groth, ``On the size of pairing-based non-interactive arguments,'' in \emph{Annual International Conference on the Theory and Applications of Cryptographic Techniques}.\hskip 1em plus 0.5em minus 0.4em\relax Springer, 2016, pp. 305--326.

\bibitem{daza2023insurance}
V.~Daza and J.~Domingo-Ferrer, ``Privacy-preserving vehicle insurance using zkps,'' \emph{arXiv preprint arXiv:2401.03118}, 2023.

\bibitem{mdpi2023vehicles}
H.~Chen and R.~Liu, ``Decentralized car data validation using zk-snarks,'' \emph{Electronics}, vol.~12, no.~18, p. 3869, 2023.

\bibitem{polygon2023whitepaper}
{Polygon Labs}, ``Polygon zkevm: Ethereum-compatible scaling with zero knowledge proofs,'' \url{https://polygon.technology/_next/static/media/polygon-zkevm-whitepaper.3d69db68.pdf}, 2023, accessed: 2025-08-09.

\bibitem{buterin2021endgame}
\BIBentryALTinterwordspacing
V.~Buterin, ``Endgame,'' \emph{Vitalik.ca}, 2021, ethereum L1 security and rollup vision. [Online]. Available: \url{https://vitalik.ca/general/2021/12/06/endgame.html}
\BIBentrySTDinterwordspacing

\bibitem{zkrollup2020benchmarking}
{Delendum Research}, ``Benchmarking zk-rollups for ethereum scaling,'' \url{https://delendum.xyz/zk-rollup-benchmarking}, 2020, accessed: 2025-08-09.

\bibitem{gaiaxCompliance}
{GAIA‑X Association}, ``{GAIA‑X Compliance Document: Openness, Security, European Sovereignty},'' \url{https://gaia‑x.eu/gaia‑x‑introduces‑the‑compliance‑document}, 2024, accessed: 2025‑08‑09.

\bibitem{gaiaxICAM}
{GAIA-X Association for Data and Cloud AISBL}, ``{GAIA-X Trust Framework and Interoperability Specifications (ICAM)},'' \url{https://gaia-x.eu/sites/default/files/2023-04/GAIA-X_TrustFramework_ICAM_v22.pdf}, 2023, accessed: 2025-08-09.

\bibitem{gaiaxSelfDesc}
{GAIA‑X Technical Committee}, ``{Self‑Description Compliance in GAIA‑X Registry and Trust Framework},'' in \emph{GAIA‑X Architecture Document}, 2022, ch. 5.5.2 Self‑Description compliance.

\bibitem{gaiaxCrossBorder}
{GAIA‑X Association}, ``{GAIA‑X and Partners to Showcase Cross‑Border Data Collaboration},'' \href{https://gaia-x.eu/newsroom/showcase-cross-border}{GAIA-X Showcase Article}, 2025, accessed: 2025‑08‑09.

\bibitem{batteryPassportGaia}
{BatteryPassport Consortium}, ``{Battery Passport Technical Guidance: GAIA‑X Trust Framework and SSI for Decentralized Identity},'' 2024, accessed: 2025‑08‑09.

\bibitem{rfc6819}
D.~Lodderstedt, M.~McGloin, and P.~Hunt, ``{OAuth 2.0 Threat Model and Security Considerations},'' \url{https://datatracker.ietf.org/doc/html/rfc6819}, 2013, accessed: 2025-08-09.

\bibitem{groth16security}
J.~Groth, ``On the size of pairing-based non-interactive arguments,'' in \emph{Annual International Conference on the Theory and Applications of Cryptographic Techniques}, 2016, pp. 305--326.

\bibitem{snarkpack}
D.~Boneh, B.~Buenz, B.~Fisch, and A.~Szepieniec, ``Snarkpack: How to aggregate snarks efficiently,'' \url{https://research.protocol.ai/blog/2021/snarkpack-how-to-aggregate-snarks-efficiently/}, 2021, accessed: 2025-08-09.

\bibitem{genes2025}
J.~L. et~al., ``{GENES: An Efficient Recursive zk‑SNARK Without Trusted Setup},'' \emph{Electronics}, 2025, r1CS merging for proof aggregation—secure under DLOG assumption.

\bibitem{plonky2recursive}
ZKM, ``{The Plonky2 Recursive Zero-Knowledge Proof},'' \url{https://www.zkm.io/blog/the-plonky2-recursive-zero-knowledge-proof}, 2025, recursive ZK proof composition for efficient verification.

\bibitem{euBatteryPassport}
EUROMOT, ``{Industry Insights into EU Battery Regulation 2023/1542},'' \url{https://www.ul.com/insights/industry-insights-eu-battery-regulation-20231542}, 2025, battery passport CE marking and QR code requirements from Feb 2027.

\bibitem{zkproofLocation2025}
D.~B. et~al., ``{Zero‑Knowledge Proof‑of‑Location Protocols for Vehicle Subsidies and Taxation Compliance},'' \emph{arXiv preprint}, 2025, privacy-preserving territorial compliance using ZKPs.

\end{thebibliography}

\end{document}